\documentclass[11 pt,a4paper,twoside]{article}
\usepackage[utf8]{inputenc}
\usepackage{amsmath}
\numberwithin{equation}{section}
\counterwithin*{figure}{section}
\usepackage{amsfonts}

\usepackage{amssymb}
\usepackage{geometry}
\usepackage{pgfplots}
\usepackage{tikz}
\usepackage{subcaption}
\usetikzlibrary{arrows.meta}
\usepackage{multirow}
\usepackage{amsmath}
\usepgfplotslibrary{fillbetween}
\usetikzlibrary{patterns}
\usepackage{pdflscape}
\usepackage{changepage}
\usepackage{indentfirst}
\usetikzlibrary{positioning, decorations.text}
\usepackage{amsmath}
\usepackage{booktabs}
\usepackage{pgfplotstable}
\usepackage[english]{babel}
\usepackage{amsthm}
\usepackage{changepage}
\usepackage{fancyhdr}

\usepackage{xcolor}
\usepackage{wrapfig}
\usepackage{pgfplotstable}
\usetikzlibrary{shapes.geometric}
\usetikzlibrary{arrows.meta}
\usepackage{glossaries}
\usepackage[normalem]{ulem}
\useunder{\uline}{\ul}{}
\usepackage{lscape}
\usepackage{mathtools}

\author{Jacques Balayla MD, MPH \footnote{To whom correspondence should be addressed: Dr. Jacques Balayla MD, MPH, CIP, FRCSC.  Obstetrician-Gynecologist. Quilligan Scholar. e-mail: jacques.balayla@mcgill.ca. Lady Davis Institute (LDI), Jewish General Hospital,  McGill University, Montreal, Quebec, Canada}}

\title{A Priori Determination of the Pretest Probability}
\date{}

\begin{document}
\maketitle  
\textbf{Abstract}. 
In this manuscript, we present various proposed methods estimate the prevalence of disease,  a critical factor in the interpretation of screening tests.  To address the limitations of these approaches, which revolve primarily around their \textit{a posteriori} nature,  we introduce a novel method to estimate the pretest probability of disease, \textit{a priori}, utilizing the Logit function from the logistic regression model.  This approach is a modification of McGee's heuristic, originally designed for estimating the posttest probability of disease.  In a patient presenting with $n_\theta$ signs or symptoms,  the minimal bound of the pretest probability, $\phi$, can be approximated by:

\begin{center}
\begin{large}
$\phi \approx  \frac{1}{5}{ln\left[\displaystyle\prod_{\theta=1}^{i}\kappa_\theta\right]}$
\end{large}
\end{center}

where $ln$ is the natural logarithm, and $\kappa_\theta$ is the likelihood ratio associated with the sign or symptom in question. 
\newpage

\section{Background}
Evidence-based medicine relies on a proper understanding of epidemiological principles and biostatistical data retrieved from well-conducted studies, which then inform the most appropriate clinical management of patients \cite{subbiah2023next}. Pretest probability is a term used in medical and diagnostic contexts to describe the likelihood or probability of a particular condition or disease being present before any diagnostic tests or investigations are conducted \cite{akobeng2007understanding}.  In Bayesian terms, it represents the prior probability for a binary classification system, such as a screening intervention that classifies individuals in one of two categories, that is,  as either ``sick" or ``not sick" \cite{van2021bayesian}.  The pretest probability, henceforth referred to as $\phi$, represents the initial estimate of the probability that a patient has a certain condition based on any signs, symptoms, and risk factors that prompt the clinician to consider specific tests in light of a potential diagnosis \cite{balayla2020prevalence}.  The pretest probability is critical because it significantly influences the interpretation of screening and diagnostic test results \cite{medow2011qualitative}.  In medical decision-making, the accuracy of a screening test is often expressed in terms of the sensitivity and specificity of a screening tool \cite{croskerry2002achieving}.  Sensitivity, or true positive rate, refers to the ability of a test to correctly identify individuals with the disease, while specificity, or true negative rate, is the ability to correctly identify individuals without the disease \cite{trevethan2017sensitivity}.  The relationship between pretest probability, sensitivity, and specificity can be understood through Bayes' theorem, which describes how the pretest probability of disease is updated based on new evidence provided by the test results \cite{bours2021bayes}.  In the screening context, the posttest probability, or the probability of having the disease after testing positive, is known as the positive predictive value, henceforth termed $\rho(\phi)$ \cite{balayla2020prevalence}.  Unfortunately,  evidence suggests that clinicians have difficulty interpreting Bayes' theorem and its influence on several of the accuracy parameters of binary classification systems \cite{whiting2015well}. Therefore, understanding and considering the pretest probability is  a critical and important step towards understanding diagnostic test results accurately and making informed clinical decisions.

\section{Pretest Probability}

As alluded to before, accounting for the pretest probability of disease is an essential prerequisite to evaluate the reliability of screening and diagnostic assessments in all areas of clinical medicine.  Despite its crucial role in the screening process, there is currently no reliable method to numerically estimate the pretest probability of disease in a given individual \cite{gayed1990formula}.  Several approaches have been proposed – but these are often population-based and retrospective in the sense that only once the total number of positive screening results are considered, can an estimate for an average pretest probability be extrapolated in hindsight \cite{gayed1990formula}.  Often, for simplicity's sake, we use the prevalence of disease as a proxy statistic, but this approach overlooks the individualized nature of risk, grouping individuals with different risk profiles together \cite{agoritsas2011does}. The consequences of this approach are several, and include the dilution of intervention effects, the inability to account for population changes in disease patterns over time, and the introduction of selection bias when selecting study participants, amongst others \cite{jackson2008dangers}.  Patients with signs and symptoms of disease have a greater pretest probability than those without,  and vice-versa \cite{sackett2002architecture}.   As a metric,  pretest probability is measured on a continuous spectrum from 0 to 100$\%$ and its value is  proportional to the positive predictive value of screening tests \cite{balayla2022bayesian}.   Though the interpretative difficulties around these concepts may in fact be a consequence of the terminology used - which isn't immediately intuitive,  it also stands to reason that Bayes' Theorem is fundamentally mathematical and therefore somewhat abstract \cite{medow2011qualitative}.  In this manuscript, we present different proposed methods to estimate the pretest probability of disease - a cornerstone of the diagnostic process.  In order to overcome their limitations, we propose a method to estimate - $\textit{a priori}$  - the pretest probability of disease using the logit function, as a modification of McGee's heuristic,  originally developed for the estimation of the posttest probability \cite{mcgee2002simplifying}. 

\subsection{Positive Likelihood Ratio ($\kappa$)}
The positive likelihood ratio (LR+) for a dichotomous test is defined as the likelihood of a test result in patients with the
disease divided by the likelihood of the test result in patients without the disease.  Otherwise stated, the positive likelihood ratio $\kappa$ gives the change in the odds of having a diagnosis in patients with a positive test \cite{deeks2004diagnostic}. For example, a $\kappa$ close to 1 means that the test result does not change the likelihood of disease or the outcome of interest appreciably. The more the likelihood ratio for a positive test  is greater than 1, the more likely the disease or outcome \cite{mcgee2002simplifying}.  We can visualise the impact of the likelihood ratio $\kappa$ on the PPV by using a Fagan nomogram \cite{caraguel2013two}.  A Fagan nomogram is a graphical tool used in medicine and clinical decision-making to calculate and illustrate the post-test probability of a medical condition based on the results of diagnostic tests and the pre-test probability of that condition.  It's named after the British epidemiologist and statistician M. J. Fagan, who developed this visualization method.  The Fagan nomogram consists of a simple chart or diagram that allows healthcare professionals to estimate how much a diagnostic test result should change their suspicion of a particular diagnosis \cite{caraguel2013two}.  It's especially useful when dealing with medical tests that have both false positive and false negative results, as it helps clinicians interpret the test results in the context of the patient's clinical presentation. To use a Fagan nomogram, you draw a straight line connecting the pre-test probability on one side to the likelihood ratio (LR) associated with the test result on the other side.  The point where the line intersects the post-test probability scale represents the updated estimate of the likelihood that the patient has the condition after considering the test result.  That the separation between the ticks on the pre and post-test probability are not equally spaced suggests that indeed the relationship between these two variables - as stated in the screening equation - are not linear.  Fagan nomograms are valuable tools because they provide a visual and intuitive way for healthcare professionals to integrate diagnostic test results into their clinical decision-making process. They help clinicians determine how much confidence they should place in a test result and whether further testing or clinical interventions are necessary.  

\section{Bayesian fundamentals in the context of screening}
Bayes' theorem provides a mathematical framework to explain how existing beliefs change in light of new evidence \cite{balayla2020prevalence}.  From Bayes' theorem, we can derive the positive predictive value equation of a screening test,   $\rho(\phi)$,  defined as the percentage of positive tests which correctly identify affected individuals.  The positive predictive value (PPV) is a function of a test's sensitivity and specificity,  but also of the pretest probability:
\begin{large}
\begin{equation}
\rho(\phi) = \frac{a\phi}{a\phi+(1-b)(1-\phi)} 
\end{equation}
\end{large} 
\\
\
where $\rho(\phi)$ = PPV,  a = sensitivity,  b = specificity and $\phi$ = prevalence/pretest probability.  The PPV, $\rho(\phi)$, is therefore a function of the pretest probability, $\phi$.  
\\
\

We have previously defined the prevalence threshold as the prevalence level in the PPV-prevalence curve below which binary classification systems are less reliable \cite{chicco2023statistical}.  In technical terms, this point represents the inflection point of maximum curvature in the aforementioned curve below which the the rate of change of a tool's positive predictive value drops at a differential pace relative to the pretest probability \cite{balayla2020prevalence},  \cite{balayla2022bayesian}.  Below this point, type I errors -  the mistaken rejection of a true null hypothesis, what is colloquially known as ``false positives" - increase \cite{memari2021low}.  This value, termed $\phi_e$, is a function of a test's sensitivity $a$ and specificity $b$,  and is defined on the pretest probability axis \cite{balayla2021formalism}:
\

\begin{large}
\begin{equation}
\phi_e =\frac{\sqrt{1-b}}{\sqrt{a}+\sqrt{1-b}}
\end{equation}
\end{large}

The corresponding positive predictive value at this prevalence level  is given by plotting the above equation into the positive predictive value equation \cite{safari2015part},  so we obtain:

\begin{large}
\begin{equation}
\rho(\phi_e)=\sqrt{\frac{a}{1-b}}\frac{\sqrt{1-b}}{\sqrt{a}+\sqrt{1-b}}=\sqrt{\frac{a}{1-b}}\phi_e
\end{equation}
\end{large}

Interestingly, the above expression leads to the well known formulation for the positive predictive value as  a function of prevalence and the positive likelihood ratio ($\kappa$), defined as the sensitivity $a$ over the compliment of the specificity $b$ \cite{deeks2004diagnostic}:

\begin{large}
\begin{equation}
\kappa = \frac{a}{1-b}
\end{equation}
\end{large} 

Graphically, an example of a PPV-prevalence curve with some sensitivity $a$ and specificity $b$, reveals the point of maximum curvature where the radius of curvature $R_c$ is at a minimum \cite{balayla2022information}.  The vertical dashed line is the prevalence threshold (Figure 1).
\newpage
\begin{small}
\begin{center}
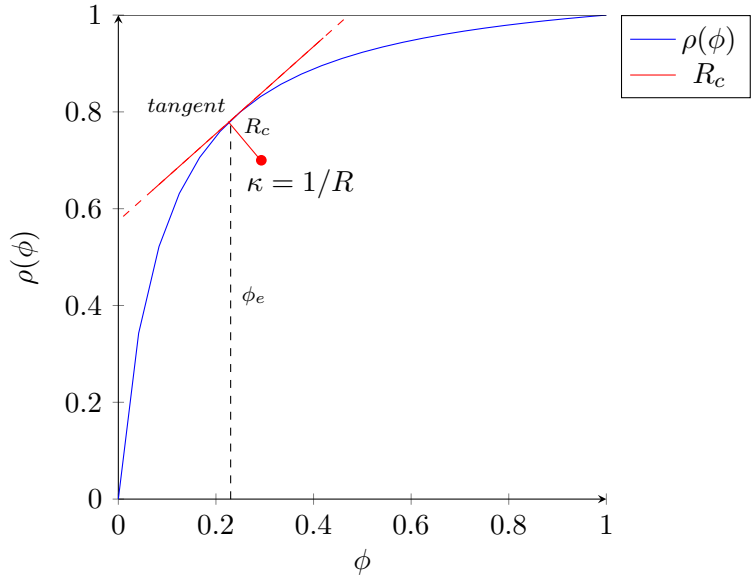
\textbf{Figure 1. PPV-pretest probability curve and the prevalence threshold $\phi_e$}
\end{center}
\end{small}
\begin{center}
\begin{tikzpicture}
	\begin{axis}[
    axis lines = left,
    xlabel = $\phi$,
	ylabel = {$\rho(\phi)$},    
     ymin=0, ymax=1,
    legend pos = outer north east,
     ymajorgrids=false,
    grid style=dashed,
    width=8cm,
    height=8cm,
     ]
	\addplot [
	domain= 0:1,
	color= blue,
	]
	{(0.60*x)/((0.60*x+(1-0.95)*(1-x))};

\addplot [	
domain=0.070:0.42, samples=100, color=red] {(0.9*x+0.575)};

\addplot [dashed,	
domain=0.01:0.62, samples=100, color=red] {(0.9*x+0.575)};

\addplot [	
domain=0.225:0.295, samples=100, color=red] {(-1.20*x+1.05005)};

\addplot+ [
domain= 0:1,	
color = black,
mark size = 0pt
 ]
	{1};
	\fill [red] (293,70) circle[radius=2pt];

	\node[above,black] at (373,60) {$\kappa=1/R$};
	\node[above,black] at (140,77) {\scriptsize{$tangent$}};
	\node[above,black] at (285,73) {\scriptsize{$R_c$}};
	\node[above,black] at (278,38) {\scriptsize{$\phi_e$}};
	\addlegendentry{$\rho(\phi)$}
	\addlegendentry{$R_c$}
	\end{axis}
\draw [dashed] (1.48,0) -- (1.48,5.0);
\end{tikzpicture}
\end{center}

Beyond this point, the rate of change of $\rho(\phi)$ is equal to $\frac{1}{\sqrt{\kappa}}$ (work not shown here), which suggests that for sufficiently adequate systems,  $\rho(\phi_e)$ is comparable to $\rho(1)$.

\subsection{Heuristics in establishing pretest probabilities}
The main theme of the theory on the prevalence threshold is that the prevalence threshold, $\phi_e$, can act as a benchmark for confidence in clinical assessments and decision-making, as the predictive values are comparable by a factor of 1/$\sqrt{\kappa}$ relative to 1 \cite{balayla2022bayesian}.  The critical question that remains though is,  how do we know where an individual lies in term of risk? How do we therefore know if we have reached the threshold or not? Indeed,  accounting for the pretest probability of disease is an essential prerequisite to evaluate the reliability of screening and diagnostic assessments in all areas of clinical medicine.  Despite its crucial role in the screening process, there is currently no reliable method to evaluate the pretest probability of disease in a given individual.  Several approaches have been proposed – but these are often population-based and retrospective in the sense that only once the total number of positive screening results are considered, can an estimate for an average pretest probability be extrapolated in hindsight.  Herein, we review some of these methods.

\subsection{Qualitative approaches to Bayes' Theorem}

The mathematical complexity of Bayes' theorem is in fact not insurmountable.  However, for clinicians,  it is also not immediately intuitive \cite{reed2013pretest}.  Several attempts to reduce the theorem's complexity have been proposed, so as to render it more available for clinical use.  Medow and Lucey's heuristic for estimating the posttest probability in individuals who undergo screening is a pragmatic tool which qualitatively categorises an individual's risk of disease prior to undergoing testing \cite{medow2011qualitative}.  In essence,  one can first categorise the pretest probability of disease as very unlikely (less likely than 10$\%$),  unlikely (10–33$\%$),  uncertain (34–66$\%$),  likely (67–90$\%$) or very likely (more likely than 90$\%$) (Table 1).  They suggest that for disorders that are very unlikely or very likely,  no further testing is needed.  On the other hand,  if the prior probability is unlikely,  uncertain or likely,  a test and a bayesian-inspired update process is incorporated to obtain a new result \cite{balayla2020derivation}.  As per the heuristic, a positive result of a good test increases the probability of the disorder by one likelihood category (eg, from uncertain to likely) and a negative test decreases the probability by one category (e.g from unlikely to very unlikely) \cite{medow2011qualitative}.  Though Medow and Lucey's heuristic reduces the mathematical abstractions of Bayes' Theorem - several potential issues arise.  Indeed,  the clinician is called to make subjective assessments about an a priori risk,  without any specific guidance as to how to proceed.  Similarly,  such strict cut-off criteria can be problematic when,  for example,  a clinician estimates a risk that hovers around two categories.  Which category do we choose? Do we proceed with testing or not? Furthermore,  a critical consideration about the screening process is seemingly overlooked.  We know the accuracy of a classification model,  such as that provided by a screening test, is dependent on several factors, notably: the classification rule or algorithm used (what is considered the disease state?),  the intrinsic characteristics of the tool used to do the classification (the sensitivity and specificity of the test), and the relative frequency of the elements being classified (prevalence of disease/pretest probability) \cite{gayed1990formula}.  And yet,  Medow and Lucey's heuristic does not account for the impact of a test's sensitivity and specificity.  Instead,  they suggest that ``using the above intuitive cut-offs,  and tests with sensitivities and specificities between 80$\%$ and 90$\%$ the above procedure is a good approximation to Bayes’ theorem" \cite{medow2011qualitative}.  This is problematic,  given that, as clinicians,  we should always strive to develop tests whose sensitivities and specificities approach 100$\%$ so as to reduce Type I and Type II errors \cite{sedgwick2014pitfalls},  and enhance the screening accuracy for a given pretest probability.  In essence,  as screening tests are refined and become more powerful over time,  Medow and Lucey's heuristic's utility is poised to decrease.  Van den Ende et al.  showcase a logarithmic scale derivation to propose a similar category-based approach of ``clinical power classes" (Table 2) towards the estimation of posttest probabilities \cite{van2007bridging}.  However,  neither method provides a formal, a priori, way to determine the pretest probability of disease.  Indeed,  Van den Ende et al.  aptly claim: ``of course, this model gives no solution for erroneous estimations of pre-test probability and LRs of tests, which would have the same deleterious effect in the classical model" \cite{van2007bridging}.  It is thus critical the we  develop methods that can better estimate the pretest probability of disease.
\newpage
\begin{center}
\textbf{Table 1.  Medow and Lucey's heuristic for pretest probability}
\end{center}
\begin{table}[h!]
\centering
\begin{tabular}{|c|c|ll} 
\cline{1-2}
\textbf{Categorical probability} & \textbf{Numerical probability} &  &   \\ 
\cline{1-2}
Very unlikely           & Less likely than 10\% &  &   \\ 
\cline{1-2}
Unlikely                & Between 10\% and 33\% &  &   \\ 
\cline{1-2}
Uncertain               & Between 34\% and 66\% &  &   \\ 
\cline{1-2}
Likely                  & Between 67\% and 90\% &  &   \\ 
\cline{1-2}
Very likely             & More likely than 90\% &  &   \\
\cline{1-2}
\end{tabular}
\end{table}

\begin{center}
\textbf{Table 2.  Van der Ende et al.  clinical power classes}
\end{center}
\begin{table}[h!]
\centering
\begin{tabular}{|cccll|}
\hline
\multicolumn{5}{|l|}{\textbf{Table 2.}}                                                               \\ \hline
\multicolumn{1}{|c|}{\textbf{$\kappa$}} & \multicolumn{1}{c|}{\textbf{$log_{10}\kappa$}} & \multicolumn{3}{c|}{\textbf{Clinical power class}} \\ \hline
\multicolumn{1}{|c|}{100}  & \multicolumn{1}{c|}{2}    & \multicolumn{3}{c|}{Very strong confirmer} \\ \hline
\multicolumn{1}{|c|}{33}   & \multicolumn{1}{c|}{1.5}  & \multicolumn{3}{c|}{Strong confirmer}      \\ \hline
\multicolumn{1}{|c|}{10}   & \multicolumn{1}{c|}{1}    & \multicolumn{3}{c|}{Good confirmer}        \\ \hline
\multicolumn{1}{|c|}{3}    & \multicolumn{1}{c|}{0.5}  & \multicolumn{3}{c|}{Weak confirmer}        \\ \hline
\multicolumn{1}{|c|}{1}    & \multicolumn{1}{c|}{0}    & \multicolumn{3}{c|}{Useless}               \\ \hline
\multicolumn{1}{|c|}{0.3}  & \multicolumn{1}{c|}{-0.5} & \multicolumn{3}{c|}{Weak excluder}         \\ \hline
\multicolumn{1}{|c|}{0.1}  & \multicolumn{1}{c|}{-1}   & \multicolumn{3}{c|}{Good excluder}         \\ \hline
\multicolumn{1}{|c|}{0.03} & \multicolumn{1}{c|}{-1.5} & \multicolumn{3}{c|}{Strong excluder}       \\ \hline
\multicolumn{1}{|c|}{0.01} & \multicolumn{1}{c|}{-2}   & \multicolumn{3}{c|}{Very strong excluder}  \\ \hline
\end{tabular}
\end{table}

\subsection{Numerical approaches}
The traditional approach in clinical medicine often draws a clear distinction between information obtained from a medical history, encompassing risk factors, exposures, and the history of the present illness, and the insights gleaned from a screening test result. The terms ``pretest" and ``posttest" conventionally mark this chronological division based on the information provided by the test \cite{puhan2005randomized}.  However, I contend that this distinction is somewhat artificial, as both sources provide valuable information about the likelihood of the presence or absence of disease.  While a screening test is typically ordered following indications from the medical history suggesting potential presence of disease, it is crucial to recognize that a screening test is not confirmatory \cite{boardman1998screening}.  Instead, it acts as an additional piece of information, akin to introducing another risk factor or symptom that influences the likelihood of a diagnosis \cite{dunn2001physician}.  Traditionally, we have been trained to consider sensitivity, specificity, and predictive values as inherent properties of a test. However, there is nothing inherently special about a test compared to other sources of information. Signs, symptoms, risk factors, and details from the medical history also possess associated sensitivity and specificity parameters \cite{devon2014sensitivity}. This understanding lies at the core foundation of ``clinical diagnoses", where a diagnosis is made in the absence of formal testing.  Each sign, symptom, exposure, risk factor, or test result, denoted as $\theta$, contributes to the overall pretest probability $\phi$.  By simple mathematical induction,  the more symptoms, the more likely the presence of disease.  As such, the pretest probability can be viewed as the cumulative effect of individual factors related to $\theta$:
\begin{center}
$\phi = \theta_1 + \theta_2 + \theta_3 + \theta_4 ....\theta_i$
\end{center}
But how does the sum of risk factors translate into probabilities ranging from 0 to 1 given that $a$ and $b$ are also parameters ranging from 0-1? Logically, each factor must have an associated weight such that their sum yields a value between 0-1. What interactions, biases and confounders are at play between different $\theta_i$? Is this sum commutative? These are important questions we must consider when estimating pretest probabilities.   In the next sections, we'll describe the ways in which pretest probabilities have been previously numerically estimated.
\subsection{Richardson's methods}
Richardson et al.  \cite{richardson1999pretest} list several approaches to estimate the pretest probability, most of which make use of previously collected data. These approaches include using disease prevalence data from clinical research studies, the prevalence data among patients presenting with the symptom/sign of interest (often derived from practice or administrative databases), the disease prevalence in the general population as obtained from population studies,  as well as the use of clinical judgement,  and clinical prediction rules/calculators if they exist.  Indeed, estimating pretest probabilities involves a combination of objective data analysis, consideration of relevant factors, and expert judgement \cite{morgan2014use}.  The specific approach will depend on the context and the nature of the problem or decision at hand.  However, as the list above suggests, we would need prevalence information to have been formally collected previously in order to make use of it - unfortunately,  this isn't always feasible nor is data always readily available.   In order to evaluate individual data points that allow for the estimation of prevalence using screening test results, we can use Marchevsky, Rogan and Gladen's method.
\subsection{Marchevsky, Rogan and Gladen's method}
Marchevsky, Rogan and Gladen introduced a method for obtaining population estimates of disease prevalence, based on known sensitivity and specificity parameters of a screening test,  as applied to $n$ individuals \cite{rogan1978estimating}, \cite{marchevsky1974errors}.  If an imperfect test is applied to a random sample of $n$ subjects, where $t$ obtain a positive result, we obtain an apparent prevalence $\hat{\phi}$ given by: 
\begin{center}
\begin{large}
$\hat{\phi}$ = $\frac{t}{n}$ = $a\phi+ (1-b)(1-\phi) = (1-b)+J\phi$
\end{large}
\end{center}
By accounting for all positives - even false positives - we say that the estimate for prevalence is ``biased" \cite{diggle2011estimating}.  This approach requires a large $n$ to be precise,  as it estimates the average prevalence of disease and less so the  individual pretest probability of participants.  Let us denote $\hat{\phi}$ as the total proportion of positive tests in a population, which is the sum of true positives and false positives over $n$ and $\phi$ is the prevalence of disease.  Then,  for the population studied, $\phi$ is given by:
\begin{equation}
\phi = \frac{\hat{\phi}-(1-b)}{a - (1-b)} = \frac{\hat{\phi}-(1-b)}{J}=  \frac{\hat{\phi}+b-1}{J}
\end{equation}
Recall that the denominator is the sum of the sensitivity and specificity - 1, which we have previously defined as Youden's $J$ statistic.  In this case, when we have a perfect test such that $a=b=1$, we would obtain $\phi$ = $\hat{\phi}$.  Because this measurement is a population estimate whose $a$ and $b$ parameters are not equal to 1 and by its nature,  refers to a sample proportion of the population,   we can apply 95$\%$ confidence intervals to this proportion estimate $\phi$ using Z-values and assuming a normally distributed estimate of $\phi$:
\begin{equation}
\phi [95\% CI] = \phi \pm 1.96 \sqrt{\frac{\phi(1-\phi)}{n}}
\end{equation}
The Rogan–Gladen method is a frequentist approach and, by assuming the prevalence as a fixed parameter, is based on classical statistical principles. This estimator shows two important  limitations.  First, if the apparent prevalence is lower than the false positive rate $(1-b)$, the Rogan–Gladen estimate will become negative, which is of course meaningless as a value of prevalence \cite{pithua2012estimated}.  Likewise,  it is not a given that the test's $a$ and $b$ parameters are known with certainty.  As such, they may be subject to uncertainty in of in themselves \cite{larremore2020jointly}.  Rogan-Gladen provide an estimate for the error in the calculation of prevalence by accounting for changes in the presumed sensitivity and specificity parameters.  However, in order to address this uncertainty more optimally,  we make use of bayesian methods.

\subsection{Revisiting Bayes' Theorem}
Recall the general statement for Bayes' theorem \cite{moons1997limitations}:
\begin{large}
\begin{equation}
P(A|B) = \frac{P(B|A) \cdot P(A)}{P(B)}
\end{equation}
\end{large}
where:
\\
\

\( P(A|B) \) is the posterior probability of event $A$ given that event $B$ has occurred,  \(P(B|A) \) is the likelihood of event $B$ given that event $A$ has occurred, \( P(A) \) is the prior probability of event $A$, and \( P(B) \) is the probability of event $B$, or marginal probability. Now, the equation above relays the form of the theorem that applies to point estimates and fixed parameters.  When the point estimates are replaced by variables, these are interpreted as probabilities and carry uncertainty about their distributions, so the theorem becomes \cite{diggle2011estimating}:
\begin{large}
\begin{equation}
f(\theta | x) = \frac{f(x | \theta) \cdot f(\theta)}{\int f(x | \theta) \cdot f(\theta) \, d\theta} \leftrightarrow f(\theta | x) \propto f(x | \theta) \cdot f(\theta)
\end{equation}
\end{large}
where:
\\
\

$f(\theta | x)$ is the posterior distribution of the parameter $\theta$ given the observed data $x$,  $f(x | \theta)$ is the likelihood function, describing the probability of observing the data $x$  given the parameter $\theta$,  $f(\theta)$ is the prior distribution, representing the initial belief or knowledge about the parameter $\theta$.  The denominator is the marginal distribution and involves integrating over all possible values of $\theta$ to ensure that the posterior distribution is properly normalized.
\\
\

In essence,  as we have seen repeatedly, the core of Bayesian analysis lies in \cite{van2021bayesian}:

\begin{center}
Posterior $\propto$ prior x likelihood
\end{center}

Put generally, the goal of bayesian statistics is to represent prior uncertainty about model parameters with a probability distribution and to update this prior uncertainty with current data to produce a posterior probability distribution for the parameter that contains less uncertainty \cite{lynch2007basics} - thus bringing us closer to the truth .
\subsection{Modelling an unknown pretest probability}
The pretest probability, as the name suggests, is a probability parameter ranging from 0 to 1. When this probability is unknown, it can be modelled using a probability distribution function and its conjugate prior \cite{d2003bayesian}. This transforms the nature of the probability from a fixed parameter in the binomial distribution to treating the probability as a random variable with a distribution of its own.  One such distribution, often termed the ``probability of probabilities," is known as the beta distribution \cite{hines2015primer}. 
\\
\

Consider a standard coin. The binomial distribution deals with a probabilistic problem, answering: given the chance of tossing heads is \( p \), what is the probability of getting \( k \) heads in \( n \) tosses? This assumes assigning an exact value to the probability of heads and, consequently, to ``not heads" (or tails). In contrast, the beta distribution approaches the question from a different perspective, addressing a more statistical problem: given that \( k \) heads appeared in \( n \) tosses, what would the probability distribution for \( p \) originally need to be to yield those results? The beta distribution examines an outcome and, based on that outcome, infers what the original distribution of \( p \) must have been to obtain those results.  It is therefore an excellent tool to extrapolate the original prior or pretest probability given some proportion of positive results \cite{chen2017clarifying}. As mentioned previously,  in classical statistics, the pretest probability is often regarded as a fixed parameter, assuming a specific value. However, in bayesian statistics, the pretest probability can be treated as a random variable with a probability distribution itself. The beta distribution is well-suited for this purpose, as it is defined on the interval \([0, 1]\) and serves as the conjugate prior for the binomial distribution \cite{chen2017clarifying}.

\subsection{Binomial distribution, beta distribution and conjugate priors}
\subsubsection{Estimation of a prevalence distribution}
The binomial distribution is a discrete probability distribution that describes the number of successes in a fixed number of independent Bernoulli trials, where each trial has only two possible outcomes. See the the toss of a coin example above as reference.  The outcomes are often denoted as ``success" and ``failure" and the probability of success is denoted by $p$. The probability function  of the binomial distribution is given by:
\[
f(X = t) = {n \choose t} p^t (1-p)^{n-t}
\]

For bayesian analyses of binary classifiers such as for screening systems, the binomial distribution represents the \textit{likelihood} function and takes the form \cite{enoe2000estimation}:

\[
f(t \vert \phi) = {n \choose t} \phi^t (1-\phi)^{n-t}
\]
where ${n \choose t}$ is the binomial coefficient, representing the number of configurations $t$ can be chosen out of $n$, $n$ is the number of trials or tests undertaken, $t$ is the number of successes or positive tests, $\phi^t$ is the probability of observing $t$ positive results given a pretest probability $\phi$,  and $(1-\phi)^{n-t}$ is the probability of observing a negative result given a pretest probability $\phi$.  
\\
\

On the other hand, the beta distribution is a continuous probability distribution defined on the interval $[0, 1]$.  It is a versatile and flexible distribution commonly employed in bayesian statistics to model random variables representing proportions or probabilities, such as the pretest probability or prevalence of disease.  The beta distribution is parametrized by two shape parameters, denoted as $\alpha$ and $\beta$, both of which must be positive \cite{olea2011use}. These parameters influence the shape of the distribution, determining its skewness and tail behaviour. For bayesian analyses of binary classifiers such as for screening systems, the beta distribution represents the conjugate \textit{prior} function and takes the form:

\[
f(\phi | \alpha, \beta) = \frac{\Gamma(\alpha + \beta)}{\Gamma(\alpha) \cdot \Gamma(\beta)} \cdot \phi^{\alpha - 1} \cdot (1 - \phi)^{\beta - 1},
\]
\

where $\phi$ is the random variable, and $\Gamma$ represents the gamma function. 
\\
\

It thus follows:
\[
f(\phi\vert \alpha, \beta) \propto \phi^{\alpha-1} (1-\phi)^{\beta-1}
\]

Notably, in bayesian inference, the beta distribution serves as the conjugate prior probability distribution for the Bernoulli, negative binomial, and geometric distributions as well \cite{martinez2014trends}. This implies that starting with a beta distribution as a prior and updating it with binomial data will result in a posterior distribution that is also a beta distribution. Therefore, the beta distribution, with parameters \( (\alpha, \beta) \), proves to be an excellent tool for estimating pretest probabilities in the screening context, given that it inherently involves a binomial nature \cite{cameron1998new}. It follows that during the modelling phase, we already know the posterior probability will also be a beta distribution \cite{lynch2007basics}.  As a result, after carrying out more experiments, one can compute the posterior by simply adding the number of successes and failures to the existing parameters \( (\alpha, \beta) \), instead of multiplying the likelihood with the prior distribution. This significantly simplifies the computation \cite{baxter2020bayesian}. The mean and variance of the beta distribution are contingent on the values of \( \alpha \) and \( \beta \), allowing practitioners to tailor the distribution of the data being modeled. In this case, the first moment or expected mean value of the beta distribution is:

\begin{equation}
E[\phi] = \frac{\alpha}{\alpha+\beta}
\end{equation}
The second moment is given by:
\begin{equation}
E[\phi^2] = \frac{\alpha(\alpha+1)}{(\alpha+\beta) (\alpha+\beta+1)}
\end{equation}
From these two parameters, we can calculate the variance of the beta distribution:
\begin{equation}
 Var(\phi) = E[\phi^2] - (E[\phi])^2 =\frac{\alpha \beta}{(\alpha + \beta)^2 (\alpha + \beta + 1)}
\end{equation}
Therefore, the standard deviation of $B(\phi; \alpha, \beta)$ becomes:
\[
\sigma(B(\phi; \alpha, \beta)) = \sqrt{\frac{\alpha \beta}{(\alpha + \beta)^2 (\alpha + \beta + 1)}}
\]
Following equation 16.4,  we thus obtain: 

\[f(\phi)=f(\phi \vert t) \propto  \phi^t (1-\phi)^{n-t} \cdot \phi^{\alpha-1} (1-\phi)^{\beta-1} = \phi^{\alpha'-1} (1-\phi)^{\beta'-1}\]
\

The updated parameters of the beta distribution become \( \alpha' = \alpha + t \) and \( \beta' = \beta + n - t \). This posterior distribution provides a probability distribution for the true pretest probability \(\phi\) given the observed number of positive tests \cite{baxter2020bayesian}.  Therefore, the beta distribution is explicitly involved in representing the prior belief, and it undergoes an update following the observation of test results in order to narrow its uncertainty. 
\subsubsection{Estimation of prevalence with an imperfect test of known parameters}
Jonathan Baxter \cite{baxter2020bayesian} proposes a method for estimating prevalence with an imperfect test having sensitivity and specificity parameters each less than 1. Assume an imperfect test is applied to a random sample of $n$ subjects, where $t$ obtain a positive result. We therefore obtain an apparent prevalence $\hat{\phi}$, as in Marchevsky, Rogan, and Gladen's method, equal to the probability that a test applied to a random individual from such a population yields a positive result.  
\newpage
Once again, $\hat{\phi}$ is given by: 
\begin{center}
\begin{large}
$\hat{\phi} = \frac{t}{n}$
\end{large}
\end{center}
Suppose we know the test's sensitivity ($a$) and specificity ($b$) parameters. Denote the population prevalence by $\phi$. The probability $\hat{\phi}$ of a positive test is the probability of a positive test given the subject has the disease, plus the probability of a positive test given the subject is disease-free:
\[\hat{\phi} = a\phi + (1-b)(1-\phi) = (1-b) + J\phi\]

The probability of retrieving $t$ positive tests out of $n$ subjects undergoing testing follows a binomial distribution with the probability parameter $\hat{\phi}$:
\[f(t|\phi, n, a, b) = \binom{n}{t} \hat{\phi}^{t} (1 - \hat{\phi})^{n-t}\]

Following Bayes’ rule, $f(\phi|t, n, a, b)$ is the posterior probability of $\phi$ given the observed data:
\[f(\phi|t, n, a, b) = \frac{f(t|\phi, n, a, b) f(\phi)}{f(t|n, a, b)}\]

where $f(\phi)$ is the prior probability of $\phi$, or the prevalence of the disease, and:
\[f(t|n, a, b) = \int_{0}^{1} f(t|\phi, n, a, b) f(\phi) \, d\phi\]

If we pick a uniform prior for $\phi$ such that $f(\phi) \propto 1$, and apply $d\phi = \frac{d\hat{\phi}}{J}$, we obtain:
\begin{align*}
f(t|n, a, b) &= \frac{1}{J} \binom{n}{t} \int_{1-b}^{a} \hat{\phi}^t (1 - \hat{\phi})^{n-t} \, d\hat{\phi} \\
&= \frac{1}{J} \binom{n}{t} \left[ \int_{0}^{a} \hat{\phi}^t (1 - \hat{\phi})^{n-t} \, d\hat{\phi} - \int_{0}^{1-b} \hat{\phi}^t (1 - \hat{\phi})^{n-t} \, d\hat{\phi} \right] \\
&= \frac{1}{J} \binom{n}{t} [B(a; t + 1, n - t + 1) - B(1-b; t + 1, n - t + 1)] \\
&= \frac{1}{J} \binom{n}{t} [B(a) - B(1-b)]
\end{align*}

Recall that in Bayesian statistics, a uniform prior (also known as a non-informative or flat prior) is a type of prior probability distribution that assigns equal probability to all possible values within a specified range. It reflects a lack of prior knowledge or bias toward any particular value in that range, where $f(\phi) \propto 1$ \cite{van2006prior}. The absence of $f(\phi)$ beyond that point means that the likelihood term dominates the posterior, and the specific form of the prior is no longer explicitly present in the final expression. Therefore, this yields:
\[f(\phi|t, n, a, b) = \left[\frac{J}{B(a) - B(1-b)}\right] [(1-b) + \phi(J)]^t [1 - (1-b) - \phi(J)]^{n-t}\]
\

Or, again, as above: 
\[f(\phi|t, n, a, b) \propto [(1-b) + \phi(J)]^t [b - \phi(J)]^{n-t}\]

A summary of the above process can be seen in the flow diagram below (Figure 2):
\

\begin{center}
\fbox{%
  \begin{minipage}{0.4\linewidth}
    \begin{center}
      \begin{tikzpicture}[node distance=2cm, auto]
        \node (prior) [align=center] {Uniform prior \\ $f(\phi)$};
        \node (likelihood) [below of=prior, align=center] {Likelihood function\\ $f(t|\phi, n,a, b)$};
        \node (posterior) [below of=likelihood, align=center] {Posterior distribution \\ $f(\phi|t, n, a, b)$};
        \draw[->] (prior) -- (likelihood) node[midway, right] {Bayes' Rule};
        \draw[->] (likelihood) -- (posterior) node[midway, right] {Normalization};
      \end{tikzpicture}
    \end{center}
  \end{minipage}%
}

\end{center}
\

\begin{center}
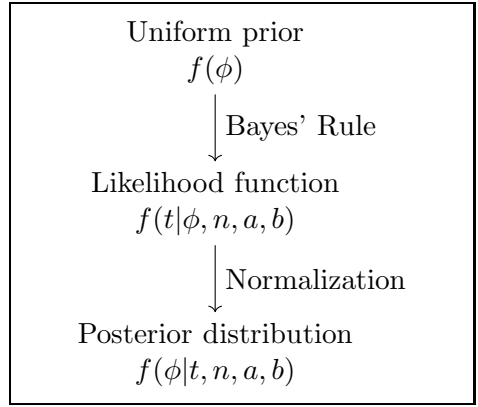
\textbf{Figure 2.  Flow diagram representing the Bayesian process to determine the average prevalence $\phi$.}
\end{center}

\subsection{Method using a test of unknown parameters}

Baxter \cite{baxter2020bayesian} as well as Larremore et al.  \cite{larremore2020jointly} propose similar techniques for scenarios in which the sensitivity and specificity parameters are unknown. If $a$ and $b$ are unknown, $\phi$ can still be estimated, albeit with reduced precision, using a Bayesian approach. This requires us to specify a prior distribution for $\phi$ and informative prior distributions for $a$ and $b$. The goal is to estimate the population $\phi$, the test's sensitivity $a$, and the test's specificity $b$, as determined by the data collected ($x$) and the validation data ($v$) used. Recall the relationship at the core of Bayes' theorem:
\begin{large}
\begin{center}
$\text{f}(\theta | x) \propto \text{f}(x | \theta) \cdot \text{f}(\theta)$
\end{center}
\end{large}

We can modify the above relationship by accounting for the unknown variables:

\begin{large}
\begin{center}
$\text{f}(\phi, a, b | x,v ) \propto \text{f}(x | \phi,a,b) \cdot \text{f}(v | a,b )$
\end{center}
\end{large}
We break down each equation to obtain the same equation as though the parameters were known, but add a new term to account for the uncertainty in the sensitivity and specificity. Suppose the test has been validated with $t_b$ false positives out of $n_b$ known negative samples, and $t_a$ true positives out of $n_a$ known positive samples. Assuming a beta prior on $b$ with parameters $\alpha_b$, $\beta_b$, the posterior density on $b$ given the validation data is proportional to a beta density with parameters $t_b + \alpha_b$, $n_b - t_b + \beta_b$. Reducing redundancy in the notation, we write $B_b(b)$ for this density and similarly $B_a(a)$ for the corresponding density on $a$. Let $B(1-b + J\phi)$ denote the original density at $(1-b) + J\phi$ of the beta distribution with parameters $t + 1$, $n - t + 1$.
\begin{equation}
\text{f}(\phi| t, n, t_a,  n_a,t_b, n_b) \propto \int_{0}^{1} \int_{0}^{a} \frac{1}{B(a) - B(1-b)} B((1-b)+J\phi) \cdot B_{b}(b)\cdot B_{a}(a)\cdot da \cdot db
\end{equation}

Thus far we have seen Richardson's list - which estimates prevalence from data collected in previous studies,  and Marchevky, Rogan and Gladen's method which utilises the number of positive tests in large cohorts undergoing screening and surmises a point estimate from the total number of positive results and the parameters of the test. We likewise saw the use of binomial distributions and conjugate priors which estimate the prevalence and pretest probability as a probability distribution instead of as a point estimate, in both cases where the test's parameters are either known or unknown. This method too requires a large number $n$ of tests conducted to best estimate the average prevalence of disease.  While all of these methods are interesting - they share one fundamental flaw - they are retrospective in nature; that is, they estimate an average prevalence in a population \textit{a posteriori}, after data has been collected.  It is therefore of no real use to a clinician who is faced with an individual patient in whom a pretest probability of disease exists, which may well vary largely from the average estimated prevalence.  Indeed, these methods are of more particular use in epidemiology and public health domains,  but they are less applicable to the individual patient in a clinical encounter.  We therefore need a technique which estimates - \textit{a priori} - the pretest probability of disease.

\subsection{Logit function and McGee's heuristic}
Given the differences between odds and probabilities, and the fact that technically $\kappa$'s measure a change in odds - we need to find a way that allows the clinician to estimate the change in post-test probability as a function of the $\kappa$ without having to make cumbersome calculations.  In order to do this,  given odds O and probability $p$:

\begin{center}
\begin{large}
O = $\frac{p}{1-p}$
\end{large}
\end{center}

In passing, by algebraic equivalence, it is also true therefore to say:
\begin{center}
\begin{large}
ln(O) = ln$\left(\frac{p}{1-p}\right)$
\end{large}
\end{center}
The function ln$\left(\frac{p}{1-p}\right)$ is known as the logit or log-odds function, and it forms the basis of the logistic regression \cite{mcgee2002simplifying}.  However,  in the equations above, we can figure out how probabilities change odds, when in fact we're interested in the opposite - how odds and $\kappa$'s change the probabilities.  Re-arranging the equation above we obtain:
\begin{center}
\begin{large}
O = $\frac{p}{1-p}$ $\leftrightarrow$ p = $\frac{O}{1+O}$
\end{large}
\end{center}
\newpage
And since the odds are just the exponential of the log-odds, the log-odds can also be used to obtain probability:
\begin{center}
\begin{large}
$p$ = $\frac{exp(ln(O))}{1+exp(ln(O))}$ = $\frac{e^{ln(O)}}{1+e^{ln(O)}}$ = $\frac{1}{1+e^{-ln(O)}}$
\end{large}
\end{center}
We can consider the ln(O) as the independent variable of a plot to assess how the post-test probability changes accordingly (Figure 3.):

\begin{center}
\begin{tikzpicture}
	\begin{axis}[
    axis lines = left,
    xlabel = $ln(O)$,
	ylabel = {$\rho(\phi)$},    
     ymin=0, ymax=1,
    legend pos = outer north east,
     ymajorgrids=true,
     xmajorgrids=true,
    grid style=dashed,
    width=8cm,
    height=8cm,
     ]
	\addplot [
	domain= -15:15,
	color= blue, thick
	]
	{1/(1+e^-x)};
\addplot [
	domain= -15:15,
	dashed,
	color= red,thick
	]
	{0.22 *x+0.5};
	\end{axis}
\end{tikzpicture}
\end{center}
\begin{center}
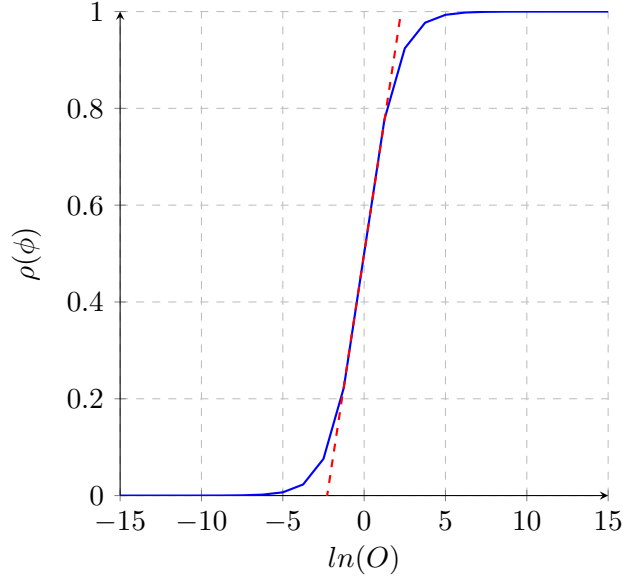
\textbf{Figure 3.  The Logit Function.}
\end{center}
A probability of 0.5 corresponds to a logit of 0.  Negative logit values indicate probabilities smaller than 0.5, and positive logits indicate probabilities greater than 0.5. The relationship is symmetrical: logits of -0.2 and 0.2 correspond to probabilities of 0.45 and 0.55, respectively. Note that the absolute distance to 0.5 is identical for both probabilities. 
This symmetry corresponds well to to the spectrum of the likelihood ratio - where below 1,  the negative predictice value $\Upsilon$ predominates, and above, $\kappa$ predominates. Indeed, for probabilities of 0.5 the odds equal 1, and the log-odds equal 0.
\\
\

In order to provide a heuristic, or shortcut approach, we can study the geometry of the logit function.  In red, a linear approximation of the logit function is observed which can act as a heuristic to determine the percentage by which the post-test probability or in this case, the positive predictive value, $\rho(\phi)$, changes in the presence of a test whose $\kappa$ or $\Upsilon$ is known.  This linear equation of the form y=mx+b is best approximated between probabilities of 10 and 90 $\%$ and is henceforth referred to as McGee's heuristic, after the author who first described it \cite{mcgee2002simplifying}:
\begin{center}
\begin{large}
$\Delta \rho(\phi) \approx$ 0.22 $\cdot$ ln($\kappa$)+ 0.5
\end{large}
\end{center}
where $\Delta \rho(\phi)$ represents the difference between the pretest and post-test probability. The 0.22 is roughly equivalent to 4.54, but we can round up to 5 for simplicity's sake.  As such, the rate of change $m$ of the linear approximation is:
\begin{center}
\begin{large}
$\boxed{m$ $\approx$ $\frac{ln(\kappa)}{5}}$ 
\end{large}
\end{center}

Most critically, notice how this change is independent of the pretest probability - that is,   whatever the starting pretest probability is the linear approximation provides the value it would increase by in the context of a positive test.  Otherwise stated:
\begin{center}
\begin{large}
$\boxed{\rho(\phi) \approx \phi+ \frac{ln(\kappa)}{5} \leftrightarrow {\rho(\phi)- \phi \approx\frac{ln(\kappa)}{5}}}$ 
\end{large}
\end{center}

With this equation, we can see why a $\kappa$ $>$ 10 is used as a standard for a ``good" test or as having ``strong" evidence for the presence of disease.  Let us assume that in an ideal situation, we want to be 100$\%$ certain of the presence of disease following a positive test, such that  $\rho(\phi)$ = 1.  What kind of $\kappa$ should such a test have? We can use the above equation by inputting $\rho(\phi)$ = 1,  to obtain:
\

\begin{center}
\begin{large}
1 - $\phi \approx  \frac{ln(\kappa)}{5}$
\end{large}
\end{center}

We obtain an equation with two variables,  $\phi$ and $\kappa$.  The table below (Table 3.) showcases what $\phi$ should be as $\kappa$ improves so as to obtain $\rho(\phi)$ = 1:

\begin{table}[h!]
\centering
\begin{tabular}{|cccc|}
\hline
\multicolumn{4}{|l|}{\textbf{Table 3. $\phi$ as a function of $\kappa$}}                                                   \\ \hline
\multicolumn{1}{|c|}{\textbf{$\kappa$}} & \multicolumn{1}{l|}{\textbf{$\frac{ln(\kappa)}{5}$}} & \multicolumn{1}{c|}{\textbf{$\phi_{1.0}$}} & $\phi_{0.5}$ \\ \hline
\multicolumn{1}{|c|}{1}  & \multicolumn{1}{c|}{0.00} & \multicolumn{1}{c|}{1.00} & 0.50 \\ \hline
\multicolumn{1}{|c|}{2}  & \multicolumn{1}{c|}{0.15} & \multicolumn{1}{c|}{0.85} & 0.35 \\ \hline
\multicolumn{1}{|c|}{3}  & \multicolumn{1}{c|}{0.24} & \multicolumn{1}{c|}{0.76} & 0.26 \\ \hline
\multicolumn{1}{|c|}{4}  & \multicolumn{1}{c|}{0.30} & \multicolumn{1}{c|}{0.70} & 0.20 \\ \hline
\multicolumn{1}{|c|}{5}  & \multicolumn{1}{c|}{0.35} & \multicolumn{1}{c|}{0.65} & 0.15 \\ \hline
\multicolumn{1}{|c|}{6}  & \multicolumn{1}{c|}{0.39} & \multicolumn{1}{c|}{0.61} & 0.11 \\ \hline
\multicolumn{1}{|c|}{7}  & \multicolumn{1}{c|}{0.43} & \multicolumn{1}{c|}{0.57} & 0.07 \\ \hline
\multicolumn{1}{|c|}{8}  & \multicolumn{1}{c|}{0.46} & \multicolumn{1}{c|}{0.54} & 0.04 \\ \hline
\multicolumn{1}{|c|}{9}  & \multicolumn{1}{c|}{0.48} & \multicolumn{1}{c|}{0.52} & 0.02 \\ \hline
\multicolumn{1}{|c|}{10} & \multicolumn{1}{c|}{0.51} & \multicolumn{1}{c|}{0.49} & 0.00 \\ \hline
\end{tabular}
\end{table}
\begin{center}
\textbf{Table 3.} $\phi$ needed to obtain $\rho(\phi)$ of 1 or 0.50 as a function of $\kappa$
\end{center}

If we define $\rho(\phi)$ as the probability of disease following a positive screening test,  then by definition, when this probability exceeds 50$\%$, the individual tested is more likely than not to have the disease in question.  Let's assume, for argument's sake,  that our prior is entirely uninformative, implying $\phi$ $\approx$ 0.  While such a scenario is unlikely as no known condition has prevalence = 0 (except but for say sex-specific conditions, such as pregnancy in a male or prostate cancer in a female), it nevertheless is important to contemplate such a situation.  In such a scenario, with $\kappa$ $>$ 10, we find that $\rho(\phi)$ $\approx$ 0.50. In other words, solely considering the test's performance, a screening test with $\kappa$ $>$ 10 indicates that, in the context of a positive result, it is more likely than not that the individual indeed has the condition in question. As depicted in the table, a $\kappa$ value exceeding 10 is the minimum required to ensure that, irrespective of the initial pretest/prior probability, the test is more likely than not a true positive, implying that the individual is more likely than not afflicted with the disease under consideration.  This fact is a simple consequence of the geometry of the logic function and may explain the reason why $\kappa$ $>$ 10 is considered to reflect strong evidence for the presence of disease - it moves a patient from less than 50$\%$ risk to beyond  50$\%$.  We can show this is the case algebraically as follows.  Taking the limit as $\phi$ goes to 0,  for a $\rho(\phi)$ = 0.5, we obtain:
\begin{center}
\begin{large}
0.50 - $0 \approx\frac{ln(\kappa)}{4.54}$ 
\end{large}
\end{center}
Isolating $ln(\kappa)$:
\begin{center}
\begin{large}
$ln(\kappa) \approx {4.54*0.5}$ 
\end{large}
\end{center}
Isolating $\kappa$ and rounding up:
\begin{center}
\begin{large}
$\kappa \approx e^{2.27} \approx 9.67 \approx 10 $ 
\end{large}
\end{center}

From the screening equation we could determine how each variable influences the other to attain a $\rho(\phi)$ = 0.50, as follows (Figure 4.):
\begin{center}
\begin{tikzpicture}
\begin{axis}[
    xlabel={$\kappa$},
    ylabel={$\phi$},
    axis lines=middle,
      ymajorgrids=true,
     xmajorgrids=true,
    grid style=dashed,
    xmin=0.01, xmax=11, 
    ymin=0, ymax=1.0, 
    samples=100, 
    domain=0.01:15, 
    legend pos= north east, 
    grid=both 
]

\addplot[blue, thick] {0.5 - ln(x)/4.5454};
\legend{$\phi \approx 0.5 - \frac{\ln(\kappa)}{5}$}

\end{axis}
\end{tikzpicture}
\end{center}

\begin{center}
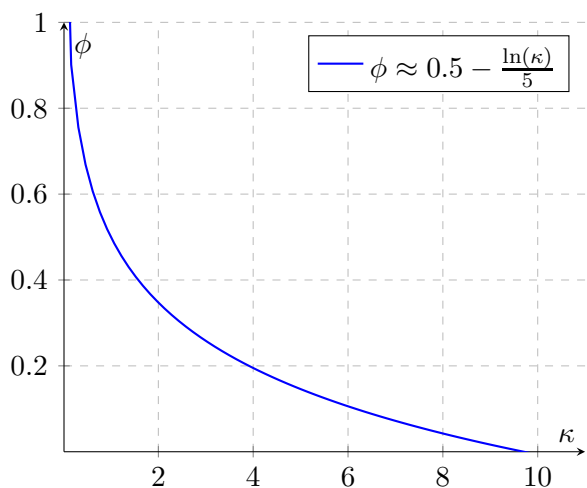
\textbf{Figure 4.}  Combination of $\phi$ and $\kappa$ to attain $\rho(\phi)$= 0.50
\end{center}

In essence,  for a given $\phi$ value,  we determine the corresponding $\kappa$ on the graph, the combination of which yields $\rho(\phi)$= 0.50 - the tipping point beyond which a positive test is more likely to be a true positive.  Heuristics like these can be very helpful to the clinical decision-making process the clinician undergoes at the bedside.  Of note, this heuristic works best for pretest probabilities between 10$\%$ and 90 $\%$. As McGee states clearly in his paper ``Although this method is inaccurate for pretest probabilities less than 10$\%$.  or greater than 90$\%$. , this is not a disadvantage, because these polar extremes of probability indicate diagnostic certainty for most clinical problems, making it unnecessary to order further tests (and apply additional LRs)" \cite{mcgee2002simplifying}.

\subsection{Heuristic approximation of a priori pretest probabilities}
At the beginning of this section,  we alluded to the use of individual risk factors $\theta_i$ to determine the overall a priori risk of disease in an individual,  $\phi$.  In essence,  the pretest probability can be viewed as the cumulative effect of individual components $\theta_i$ (such as signs, symptoms, risk factors, etc...) whose sum leads to $\phi$:
\begin{center}
$\phi = \theta_1 + \theta_2 + \theta_3 + \theta_4 ....\theta_i$
\end{center}
In this section, we'll explore intuitive ways that transform these risk factors into a probability of disease.  In particular, we'll make use of the logistic regression and the likelihood ratio.  Recall the relationship between $\kappa$,  $\rho(\phi)$ and $\phi$:
\begin{center}
\begin{large}
$\rho(\phi) =  \frac{a\phi}{ a\phi+(1-b)(1-\phi)} = \frac{\kappa\phi}{1 + ({\kappa}-1)\phi}$
\end{large}
\end{center}
In previous work, we used the logit function as dictated by McGee's heuristic \cite{mcgee2002simplifying}, to evaluate an approximation between $\kappa$,  $\rho(\phi)$ and $\phi$ for pretest probability values ranging between 0.10 and 0.90:
\begin{center}
\begin{large}
$\rho(\phi) \approx \phi+ \frac{ln(\kappa)}{5} \leftrightarrow { \phi \approx\rho(\phi)-\frac{ln(\kappa)}{5}}$ 
\end{large}
\end{center}

Let us denote the difference between $\rho(\phi)$ and $\phi$ as $\Delta p$, such that:
\begin{center}
\begin{large}
$\Delta p \approx\frac{ln(\kappa)}{5}$ 
\end{large}
\end{center}
Bridging the pretest to the posttest probability,  $\Delta p$ can be thought of as the gain in pretest probability following a positive screening test.  Now, since $\kappa$ is a function of the sensitivity $a$ and the specificity $b$,  it follows that it is prevalence independent.  We have thus associated a prevalence-independent measure,  as is the likelihood ratio $\kappa$ - to a probability of disease.  Since $\phi$ and $\rho(\phi)$ are probabilities,  they each can hold values between 0 and 1,  representing minimal and maximal bounds for both $\phi$ and $\rho(\phi)$ can take.  Let's look at both scenarios a little closer.
\subsubsection{Maximal bound of $\phi$}
On the one hand,  intuitively, we know that $\phi$ is maximally bound at 1 - by  definition.  That is, if $\phi$ = 1 then given the invariant points on the probability square plane $S$, no matter the level of $\kappa$ applied, $\rho(\phi)$ = 1. That said,  a theoretical $\phi$ = 1 would obviate the need for testing in the first place.
\begin{center}
\begin{large}
$\rho(\phi) =   \frac{\kappa(1)}{1 + ({\kappa}-1)(1)}$ = $\frac{\kappa}{\kappa} = 1$
\end{large}
\end{center}
\subsubsection{Minimal bound of $\phi$}
If we assume that the pretest probability $\phi$ is such that when subjected to a positive screening test $\rho(\phi)$ is maximized to 1,  then it would follow that $\phi$ would increase by: 
\begin{center}
\begin{large}
$\phi \approx1-\frac{ln(\kappa)}{5}$ 
\end{large}
\end{center}
Recall that when subject to orthogonal testing - the use of different ``tests" with individual sensitivity and specificity parameters (a thus $\kappa$, too) - $\rho(\phi)$ requires multiplication of the individual parameters \cite{balayla2020derivation}.  As such, the gain in pretest probability $\phi$ is given by:
\begin{center}
\begin{large}
$\phi \approx1-\frac{1}{5}{ln\left(\displaystyle\prod_{\theta=1}^{i}\kappa_\theta \right)}$ 
\end{large}
\end{center}
Therefore,  given that such equation represents the $\textit{gain}$ in $\phi$,  the the underlying pretest probability must be its complement, such that: 
\begin{center}
\begin{large}
$\phi \approx 1 -\left[1-\frac{1}{5}{ln\left(\displaystyle\prod_{\theta=1}^{i}\kappa_\theta\right)}\right] = \frac{1}{5}{ln\left[\displaystyle\prod_{\theta=1}^{i}\kappa_\theta\right]}$
\end{large}
\end{center}
We say that this is a minimum in the sense that for a given $\kappa$,  this represent the minimal pretest probability needed to obtain a $\rho(\phi)$ = 1. However, it is possible that for the same $\kappa$ a higher $\phi$ would also yield a $\rho(\phi)$ = 1.  We can illustrate graphically (Figure 5.) how these bounds change as a function of $\displaystyle\prod_{\theta=1}^{i}\kappa_\theta$, as follows:

\begin{center}
\begin{tikzpicture}
\begin{axis}
[
    xlabel={$\displaystyle\prod\kappa_\theta$}, 
    ylabel={$\phi$}, 
    grid, 
    xmin=-0, 
    xmax=100, 
    ymin=0, 
    ymax=1.0, 
     legend pos = north east,
]
     \addplot[red, thick,domain=0:100, samples=100] {0.2*ln(x)};
      \addplot[red, thick,domain=0:100, samples=100] {1};
      
\end{axis}
\end{tikzpicture}
\end{center}
\begin{center}
\textbf{Figure 5}.  Range of minimal and maximal potential values of $\phi$ based on the presence of signs, symptoms, or risk factors, each with an individual $\kappa$.
\end{center}

These values are conservative in the sense that they represent the widest possible intervals for the calculation of $\phi$ given a series of $\kappa_\theta$.  Given the heuristic nature of this technique,  it is probably best to err in the side of caution when estimating pretest probabilities by selecting the minimal bound as the approximation.  This way,  we yield the highest possible impact to the $\kappa$ of the test to bridge the pretest to the posttest probability.  Nevertheless, we could calculate an average $\phi$ as simply the mean value of both functions,  with the interval min-max range, R$(\bar{\phi})$,  as follows (Figure 6.):

\begin{center}
\begin{large}
$\bar{\phi} \approx  \frac{1+\frac{1}{5}{ln\left[\displaystyle\prod_{\theta=1}^{i}\kappa_\theta\right]}}{2}$,  R$(\bar{\phi})$ = $\left[\frac{1}{5}{ln\left(\displaystyle\prod_{\theta=1}^{i}\kappa_\theta\right)}, 1 \right]$
\end{large}
\end{center}

\begin{center}
\begin{tikzpicture}
\begin{axis}
[
    xlabel={$\displaystyle\prod\kappa_\theta$}, 
    ylabel={$\phi$}, 
    grid, 
    xmin=-0.01, 
    xmax=100, 
    ymin=0, 
    ymax=1.0, 
     legend pos = north east,
]
     \addplot[red, thick,domain=0:100, samples=100] {0.2*ln(x)};
      \addplot[red, thick,domain=0:100, samples=100] {1};
        \addplot[blue, thick,dashed,domain=-0.01:100, samples=100, name path = B] {0.5+0.1*ln(x)};
\addplot[color = black, dashed, thick] coordinates {(0, 0) (0,1)};        
\addplot[color = black, dashed, thick] coordinates {(10, 0.46) (10,1)};        
\addplot[color = black, dashed, thick] coordinates {(20, 0.599) (20,1)};
\addplot[color = black, dashed, thick] coordinates {(30, 0.68) (30,1)};
\addplot[color = black, dashed, thick] coordinates {(40, 0.74) (40,1)};
\addplot[color = black, dashed, thick] coordinates {(50, 0.78) (50,1)};
\addplot[color = black, dashed, thick] coordinates {(60, 0.82) (60,1)};
\addplot[color = black, dashed, thick] coordinates {(70, 0.85) (70,1)};
\addplot[color = black, dashed, thick] coordinates {(80, 0.876) (80,1)};
\addplot[color = black, dashed, thick] coordinates {(90, 0.899) (90,1)};
\addplot[color = black, dashed, thick] coordinates {(100, 0.92) (100,1)};
\end{axis}
\end{tikzpicture}
\end{center}
\begin{center}
\textbf{Figure 6}.  Range of minimal and maximal potential values of $\phi$ based on the presence of signs, symptoms, or risk factors, each with an individual $\kappa$. The blue dashed line represents the mean value of the pretest probability $\bar{\phi}$ as a function of $\kappa_\theta$ and the black dashed lines represent the range of potential values for $\phi$.
\end{center}

Though it appears like it,  the graph above is misleading in that there is no asymptote to that equation. In other words, for a sufficient product of $\kappa_\theta$,  the pretest probability would go beyond 1. This is not a real problem since that would imply a $\kappa_\theta$ of near 150! Recall we showed earlier that a combined $\kappa$ of 10 would be sufficient to deem a test a ``good confirmer",  so it would be unnecessary to obtain levels significantly higher than that to rule-in the presence of disease.  Similarly,  the premise of this theory is that the prevalence threshold $\phi_e$ may serve as a good alternate benchmark to rule-in disease for adequate-enough classification systems.  In the figure below,  one can see that the intersection of these two functions happens at a $\kappa$ level of $\kappa_\theta$ $\approx$ 4.5 (Figure 7).
\newpage
\begin{center}
\begin{tikzpicture}
\begin{axis}
[
    xlabel={$\displaystyle\prod\kappa_\theta$}, 
    ylabel={$\phi$}, 
    grid, 
    xmin=-0.01, 
    xmax=100, 
    ymin=0, 
    ymax=1.0, 
     legend pos = north east,
]
     \addplot[red, thick,domain=0:100, samples=100] {0.2*ln(x)};
        \addplot[blue, thick,dashed,domain=0.01:100, samples=100] {1/(x^(0.5)+1)};
\end{axis}
\end{tikzpicture}
\end{center}

\begin{center}
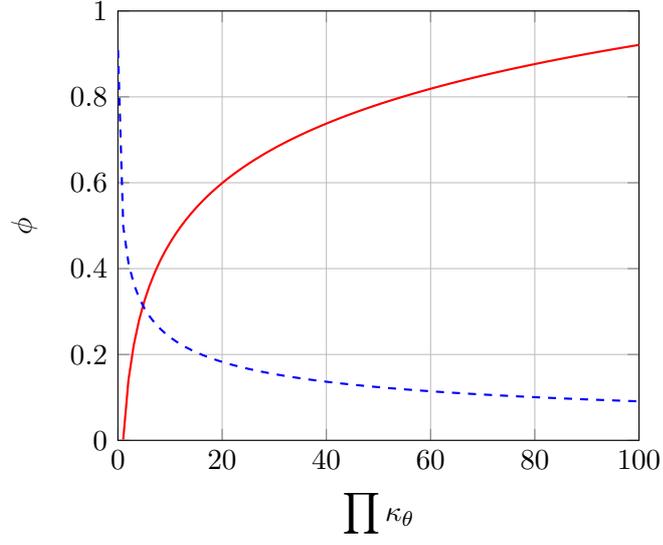
\textbf{Figure 7}.  Intersection of $\phi$ and $\phi_e$ as a function of $\kappa_\theta$ .
\end{center}

\begin{table}[h!]
\centering
\begin{tabular}{|c|c|c|c|}
\hline
\textbf{$\kappa_\theta$} & \textbf{$\bar{\phi}$} & \textbf{Min} & \textbf{Max} \\ \hline
1                   & 0.50                  & 0.00         & 1            \\ \hline
2                   & 0.57                  & 0.14         & 1            \\ \hline
3                   & 0.61                  & 0.22         & 1            \\ \hline
4                   & 0.64                  & 0.28         & 1            \\ \hline
5                   & 0.66                  & 0.32         & 1            \\ \hline
6                   & 0.68                  & 0.36         & 1            \\ \hline
7                   & 0.69                  & 0.39         & 1            \\ \hline
8                   & 0.71                  & 0.42         & 1            \\ \hline
9                   & 0.72                  & 0.44         & 1            \\ \hline
10                  & 0.73                  & 0.46         & 1            \\ \hline
20                  & 0.80                  & 0.60         & 1            \\ \hline
30                  & 0.84                  & 0.68         & 1            \\ \hline
40                  & 0.87                  & 0.74         & 1            \\ \hline
50                  & 0.89                  & 0.78         & 1            \\ \hline
60                  & 0.91                  & 0.82         & 1            \\ \hline
70                  & 0.92                  & 0.85         & 1            \\ \hline
80                  & 0.94                  & 0.88         & 1            \\ \hline
90                  & 0.95                  & 0.90         & 1            \\ \hline
100                 & 0.96                  & 0.92         & 1            \\ \hline
\end{tabular}
\end{table}
\begin{center}
\textbf{Table 4}.  Mean value of $\phi$ and its associated min-max range as a function of $\displaystyle\prod\kappa_\theta$.
\end{center}

Thus,  by accounting for the individual $\theta_\theta$, or signs and symptoms, we can multiply their $\kappa_\theta$ to approximate $\phi$ through this equation. Then - we can update this probability using Bayes' theorem with a screening test to obtain a final probability of disease.  As such:

\begin{center}
\begin{large}
$\rho(\phi) = \frac{\kappa\phi}{1 + ({\kappa}-1)\phi} \rightarrow \frac{\kappa \frac{1}{5}{ln\left[\displaystyle\prod_{\theta=1}^{i}\kappa_\theta\right]}}{1 + ({\kappa}-1)\frac{1}{5}{ln\left[\displaystyle\prod_{\theta=1}^{i}\kappa_\theta\right]}}$ 
\end{large}
\end{center}

In deriving the equation for $\bar{\phi}$ we made the assumption that we wanted to obtain a perfect positive predictive value,  such that $\rho(\phi)$ = 1.  In reality, this is likely not necessary. What would the effect of considering a different threshold for $\rho(\phi)$ be? 

\begin{center}
\begin{tikzpicture}
\begin{axis}
[
    xlabel={$\displaystyle\prod\kappa_\theta$}, 
    ylabel={$\phi$}, 
    grid, 
    xmin=-0, 
    xmax=100, 
    ymin=0, 
    ymax=1.0, 
     legend pos = north east,
]
     \addplot[red, thick,domain=0:100, samples=100] {0.2*ln(x)};
     \addplot[blue, thick,domain=0:100, samples=100] {0.1+0.2*ln(x)};
     \addplot[orange, thick,domain=0:100, samples=100] {0.2+0.2*ln(x)};
     \addplot[black, thick,domain=0:100, samples=100] {0.3+0.2*ln(x)};
      \addplot[red, thick,domain=0:100, samples=100] {1};
      
\end{axis}
\end{tikzpicture}
\end{center}

\begin{center}
\textbf{Figure 8}.  Range of minimal and maximal potential values of $\phi$ based on the presence of signs, symptoms, or risk factors, each with an individual $\kappa$,  assuming the ultimate target would be a $\rho(\phi)$ of 1.00 (red), 0.90 (blue), 0.80 (orange),  0.70 (black). 
\end{center}

We see thus in Figure 8. that lowering the ultimate $\rho(\phi)$ threshold narrows the range of $\bar{\phi}$.  Since we seek ultimately to err in the side of caution using this heuristic, we're better off using $\rho(\phi)$ = 1 as the most appropriate approach, as it yields the widest range of values, and thus the lowest bound of $\bar{\phi}$.
\newpage

\subsection{The effect of baseline prevalence on $\kappa_\theta$}

While the product of individual $\kappa_\theta$ is a good way to approximate the pretest probability of disease - a criticism could be made that the baseline risk - or prevalence of disease - is not accounted for in this calculation.  The question whether $\kappa_\theta$ accounts for the prevalence of disease is to some degree philosophical and academic.  But let us for argument's sake assume that prevalence is not accounted for in this calculation.    The effect of accounting for baseline prevalence would be simply to narrow the range of $\bar{\phi}$ by moving the line upwards. Since we err in the side of caution by taking the widest possible range - this addition would not modify the estimates for the purposes of this heuristic.   Secondly, in the presence of a high enough $\kappa_\theta$, the impact of baseline prevalence would likely be negligible for the simple reason that if the prevalence was high enough we would not need to account for $\kappa_\theta$ in the first place, as the individual would likely be at a risk level such that $\epsilon >$ $\phi_e$,  where $\epsilon$ is the prevalence of disease.

\begin{center}
\begin{large}
$\phi \approx  \frac{1}{5}{ln\left[\displaystyle\prod_{\theta=1}^{i}\kappa_\theta\right]} + \epsilon$
\end{large}
\end{center}

As we have thus seen, using the logit function and the rules of logistic regression allowed us to approximate the impact of the likelihood ratios of individual signs and symptoms. These are then used to approximate - a priori - the lower bound of the pretest probability of disease.  Clinical confirmation of these relationships should be undertaken prior to clinical implementation. 

\newpage
\bibliographystyle{unsrt}
\bibliography{references}

\end{document}